\documentclass[twocolumn]{aastex631}
\usepackage{units}
\shorttitle{Ellipsars}
\shortauthors{DuPont et al.}

\graphicspath{{./}{figures/}}
\graphicspath{{./}{figs/}}
\begin{document}

\title{Ellipsars: Ring-like explosions from flattened stars}

\correspondingauthor{Marcus DuPont}
\email{md4469@nyu.edu}

\author[0000-0003-3356-880X]{Marcus DuPont}
\author[0000-0002-0106-9013]{Andrew MacFadyen}
\affiliation{Center for Cosmology and Particle Physics, Department of Physics\\
New York University \\
New York, NY 10003, USA}

\author[0000-0002-1895-6516]{Jonathan Zrake}
\affiliation{Department of Physics and Astronomy\\
Clemson University \\ 
Clemson, SC, 29634, USA}



\newcommand{\am}[1]{{\textcolor{red} {\bf #1}}} 
\newcommand{\jz}[1]{{\textcolor{orange} {\bf #1}}} 
\newcommand{\rmtext}[1]{\textcolor{red}{\sout{\bf #1}}}
\newcommand{\marc}[1]{\textcolor{violet}{#1}} 
\newcommand{\mrev}[1]{\textcolor{blue}{\bf #1}} 
\newcommand{\newtext}[1]{\textcolor{red}{\bf #1}}
\newcommand{\adgam}{\hat{\gamma}}
\newcommand{\kms}{km s$^{-1}$}
\newcommand{\gcc}{g cm$^{-3}$}
\newcommand{\rsun}{$R_\odot$}
\newcommand{\ergcc}{erg cm$^{-3}$}
\newcommand{\enorm}{$E_{51}$}
\newcommand{\gb}{$\Gamma \beta$}
\newcommand{\eps}{$\varepsilon$}
\newcommand{\msun}{$M_\odot$}

\makeatletter
\def\convertto#1#2{\strip@pt\dimexpr #2*65536/\number\dimexpr 1#1}
\makeatother

\begin{abstract}
The stellar cataclysms producing astronomical transients have long been modeled as either a point-like explosion or jet-like engine ignited at the center of a spherically symmetric star. However, many stars are observed, or are expected on theoretical grounds, not to be precisely spherically symmetric, but rather to have a slightly flattened geometry similar to that of an oblate spheroid. Here we present axisymmetric two-dimensional hydrodynamical simulations of the dynamics of point-like explosions initiated at the center of an aspherical massive star with a range of oblateness.  We refer to these exploding aspherical stars as ``ellipsars'' in reference to the elliptical shape of the iso-density contours of their progenitors in the two-dimensional axisymmetric case. We find that ellipsars are capable of accelerating expanding rings of relativistic ejecta which may lead to the production of astronomical transients including low-luminosity GRBs, relativistic supernovae, and Fast Blue Optical Transients (FBOTs.)


\end{abstract}



\section{Introduction} \label{sec:intro}

Astrophysical transients are a topic of intense current interest, especially due to multiple time-domain surveys of the sky currently planned or in progress \citep[e.g.,][]{Barthelmy-2005,Shappee+2014, Chambers+2016,Kochanek+2017, Ivezi+2019, Bellm+2019}. These surveys are revealing an exciting variety of new transients, including low-luminosity gamma-ray bursts (llGRBs), relativistic supernovae (SNe) and Fast Blue Optical Transients (FBOTs). These phenomena have been interpreted as new types of stellar explosions requiring the ejection of a fraction of the star at relativistic or near-relativistic speed. 

Explosions in massive stars have been studied for decades \citep[e.g.,][]{Baade+Zwicky-1934, Fowler+Hoyle-1964, Filippenko-1997, Woosley-2002, Woosley+Bloom-2006, Janka-2007, Gal-Yam-2019}, though much of this research has considered spherical explosions in spherical stars, which typically produce little or no relativistic ejecta. However, observations of supernovae (SNe) associated with high-energy transients (GRBs, X-ray Flashes) often indicate the presence of relativistic outflow.  Most explanations of the relativistic ejecta component involve jetted explosions in spherical stars \citep[e.g.,][]{MacFadyen-1999,Aloy+2000,MacFadyen+2001,Zhang+2003,Wheeler+2000,Tchekhovskoy+2008,Bromberg+2011,Moesta+2014} with recent studies exploring the dynamics of the ensuing oblique shock breakout from spherical stars \citep[e.g.,][]{Matzner+2013,Salbi+2014,Afsariardchi+2018, Irwin+2021}.

Jetted explosions require the extraction of rotational kinetic energy from a rotating stellar core, implying that the progenitor stars in which these explosions take place must be at least slightly aspherical.
Furthermore, some massive stars have been directly observed to have envelopes rotating near breakup \citep[e.g.,][]{Porter+2003}, some exhibiting oblateness parameters as large as 0.5 \citep[][]{Kervell+2006,Carciofi+2008}.
Additionally, surveys of massive stars indicate that 70\% are in close binary systems \citep[][]{Sana+et+al+2012}.\@ These observations give rise to the possibility that many or most massive stars have tidally distorted envelopes.

It is thus reasonable on both observational and theoretical grounds to consider point-like explosions in aspherical stars, and to investigate the degree to which the explosion dynamics might be capable of producing relativistic ejecta. In this Letter, we present a suite of two-dimensional axisymmetric simulations of point explosions in an $18 M_\odot$ pre-supernova helium star, which is flattened to varying degrees of oblateness. Due to the elliptical shape of the iso-density contours in the pre-explosion star, we dub these explosions ``ellipsars.'' 

Our Letter is organized as follows: Section \ref{sec: form} describes the numerical setup and initial conditions; in Section \ref{sec: results} we present our results, and Section \ref{sec: discussion} discusses the astrophysical implications and relevance of our work. 

\section{Numerical setup} \label{sec: form}
\subsection{Equations of motion}\label{sec: numerics}
Our simulations of point-like explosions of oblate stars are based on time-dependent solutions of the relativistic hydrodynamics equations, in 2D spherical-polar coordinates ($r$--$\theta$) with axial symmetry ($\partial/\partial{\phi} = 0$). In an explicitly conservative form, the equations of motion are the mass-continuity equation: $\partial_\mu(\rho u^\mu ) = 0$, and the conservation of energy and momentum: $\partial_\mu(T^{\mu \nu}) = 0$, where $T^{\mu \nu}= \rho h u^\mu u^\nu + p \eta^{\mu \nu}$ is the stress-energy tensor for a perfect fluid.

In these equations, $\rho$ is the proper (i.e. comoving) mass density, $h = 1 + \epsilon + p/\rho$ is the specific enthalpy, $p$ is the gas pressure, $\epsilon$ is the specific internal energy, $\adgam = 4/3$ is the adiabatic index, $\eta^{\mu \nu}$ is the Minkowski metric with signature $(- + + +)$, and $u^\mu$ is the fluid four-velocity.
The system is closed by the adiabatic equation of state $p = (\adgam - 1)\rho \epsilon$.

Numerical solutions are obtained using a standard second-order Godunov method \cite[see e.g.][]{Marti+Muller-2003, Marti+Muller-2015}. Simulations use the generalized minmod slope limiter for piecewise linear reconstruction, with the least diffusive $\theta$-parameter, $\theta = 2$ and we invoke the method described in \citet[]{Quirk+1994} to suppress grid-aligned shock instabilities in the flow.\@ Results are presented for simulations with 5609 radial zones and 4096 angular zones, for which our results are numerically well-converged.\@ The equations are implemented in a new GPU-accelerated hydrodynamics code, \texttt{SIMBI}, which was written by this study's lead author, and is publicly available on GitHub. The code is written in a combination of Python and C++ (with CUDA or ROCm language extensions), and supports execution on multi-core CPUs (parallelized with OpenMP), and on NVIDIA and AMD GPU hardware.

\subsection{Initial Conditions}\label{sec: init_conditions}
\begin{deluxetable}{lll}[ht!]
\tablenum{1}
\tablecaption{Stellar Model Parameters\label{tab:constants}}
\tablewidth{0pt}
\tablehead{
Variable & Definition & Value\\
}
\startdata
$\rho_\odot$ & Characteristic density scale & $3 M_\odot / (4 \pi R_\odot^3)$ \\
$p_\odot$    & Characteristic pressure scale & $3 M_\odot c^2 / (4 \pi R_\odot^3)$ \\
$t_\odot$& Characteristic time scale & $R_\odot / c$ \\
$\rho_c$ & Central density & $3 \times 10^7 \rho_\odot$  \\
$\rho_{\rm wind}$ & Wind density at surface & $10^{-9} \rho_\odot$ \\
$v_{\rm{wind}}$ & Terminal speed of stellar winds & $10^3$ \kms \\
$R_1$ & First break radius & 0.0017 $R_\odot$ \\
$R_2$ & Second break radius & 0.0125 $R_\odot$ \\
$R_3$  & Spherical outer radius & 0.65 $R_\odot$\\
$k_1$  & First break slope & 3.25\\
$k_2$  & Second break slope & 2.57 \\
$n$      & Atmospheric cutoff slope & 16.7 \\
\enddata
\end{deluxetable}

Our initial conditions include an oblate progenitor star with a surrounding wind-like medium, and a spherical high-pressure region at the center to initiate a point-like explosion. We use the MESA code \citep{Paxton-2011, Paxton-2013} to evolve a star before collapse, where the initial stellar model was a low metallicity $30 M_{\odot}$ main sequence star, rotating near 99\% of its breakup velocity. By the time the star explodes, it is an $18 M_{\odot}$ Wolf-Rayet star with about half the radius of the sun. The stellar evolution model is 1D, yielding a spherically symmetric density profile for the stellar envelope. This density profile could be tabulated and accessed by the code to generate an initial condition, however the simple analytic fitting formula,
\begin{equation}
    \rho_{\rm 1D}(r) = \frac{\rho_c \times \max \left(1 - r / R_3, 0\right)^n}{1 + (r / R_1)^{k_1} / \left[1 + (r / R_2)^{k_2}\right]}
    \label{eq: 1d_density_profile}
\end{equation}
provides a close approximation to the density structure in the MESA model \cite[see][]{Duffell2015}, and is more convenient for code implementation. Here, $\rho_c$ is central density, $R_3$ is the stellar radius, $R_{1,2}$ are the first and second break radii of the density profile, and $k_{1,2}$ are the power-law slopes of the density profile between the breaks. Numerical values of these parameters are given in Table \ref{tab:constants}.

To generate a simple model for oblate stellar progenitors, the spherical density iso-surfaces at radius $r$ are mapped to oblate ellipsoids. This is accomplished by replacing $R_3$ in Equation \ref{eq: 1d_density_profile} with
\begin{equation}
    \label{eq: ellipse}
    R_\varepsilon(\theta) = \frac{a b}{\sqrt{(a\cos\theta)^2 + (b\sin\theta)^2}}; \quad \varepsilon \equiv 1 - b/a \, ,
\end{equation}
where $\varepsilon \in [0, 1]$ is the flattening parameter.
%
%
The semi-major and semi-minor axes $a$ and $b$ are determined from the condition that the progentitor volume $\propto a^2 b$ is independent of $\varepsilon$, and they satisfy $a = R_3(1 - \varepsilon)^{-1/3}$ and $b = R_3(1 - \varepsilon)^{2/3}$. When $\varepsilon=0$, the star is spherical, $R_{\varepsilon = 0}(\theta) = R_3$.

To model the circumstellar environment, we place the wind-like density profile
\begin{equation}
    \rho_{\rm wind}(r) = \frac{\dot M}{4\pi r^2 v_{\rm wind}} = \frac{A}{r^2}
\end{equation}
outside the stellar envelope, where $r > R_\varepsilon(\theta)$. The parameter $A$ is conventionally non-dimensionalized as $A = A_* \times \unit[5 \times 10^{11}]{g \ cm^{-1}}$. In our simulations, we adopt the fiducial value $A_* = 0.1$, motivated by values calculated for sub-solar metallicity ($Z=0.1Z_\odot$) Wolf-Rayet stars \cite[e.g.][]{Vink+deKoter+2005, Vink+2021}. The wind speed is set to $v_{\rm wind} = \unit[10^3]{km \ s^{-1}}$, which for $A_* = 0.1$ corresponds to a mass lass rate of $10^{-6} M_\odot \rm yr^{-1}$.

We initiate the explosion by placing a high-pressure region at small radii $r < r_{\rm exp}$,
\begin{equation}
    p(r) = \frac{3(\adgam - 1) E_{\rm exp}}{4 \pi r_{\rm exp}^3} \times H(r - r_{\rm exp}) \, ,
\end{equation}
where $H(r)$ is the Heaviside step function and $E_{\rm exp}$ is the explosion energy. Otherwise, the gas in the progenitor star and the wind medium is assumed to be cold, with $p / \rho$ set to a small number around $10^{-6}$.
\begin{figure}
\centering
\gridline{\fig{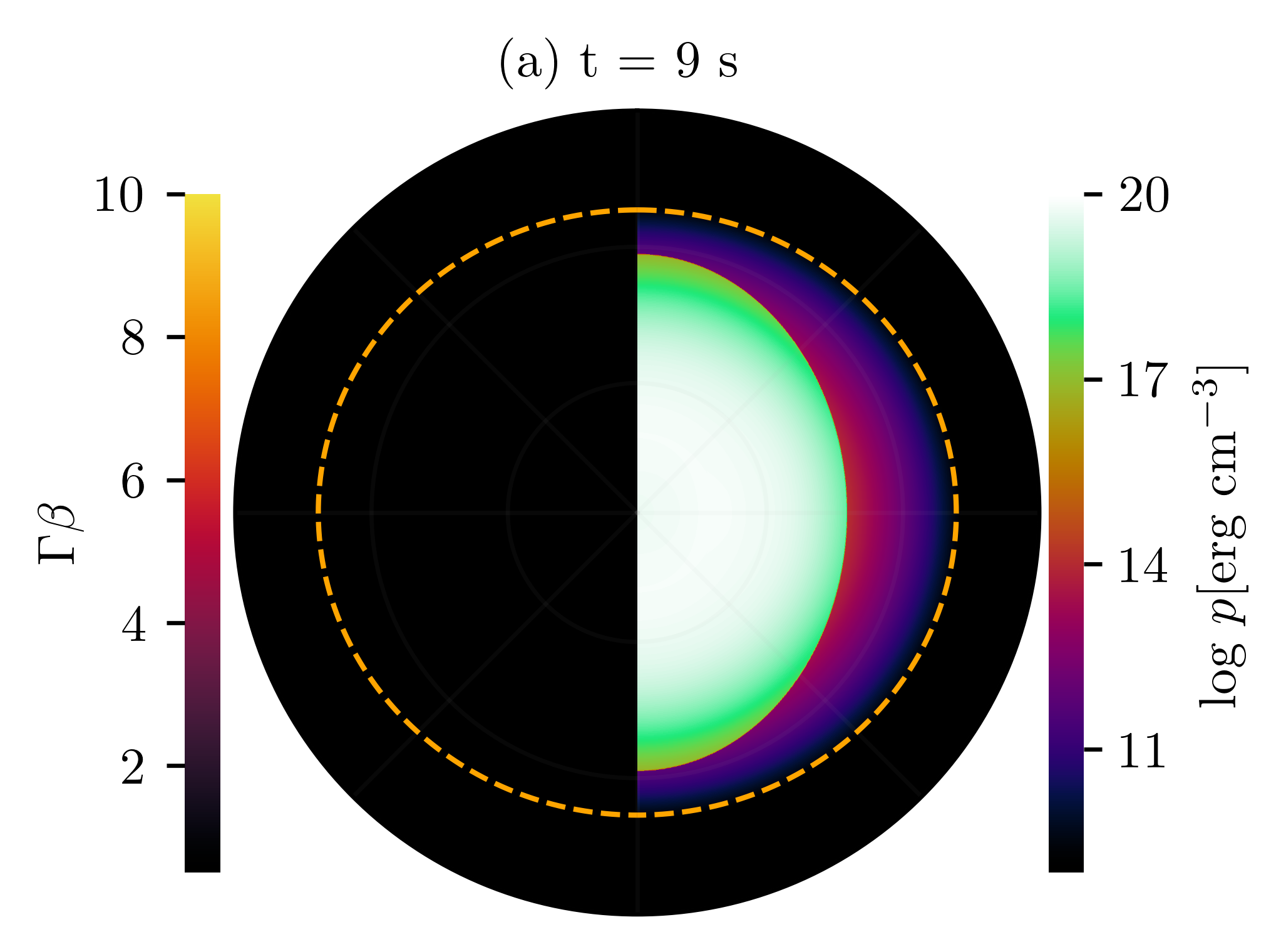}{\columnwidth}{}}\vspace*{-3.5em}
\gridline{\fig{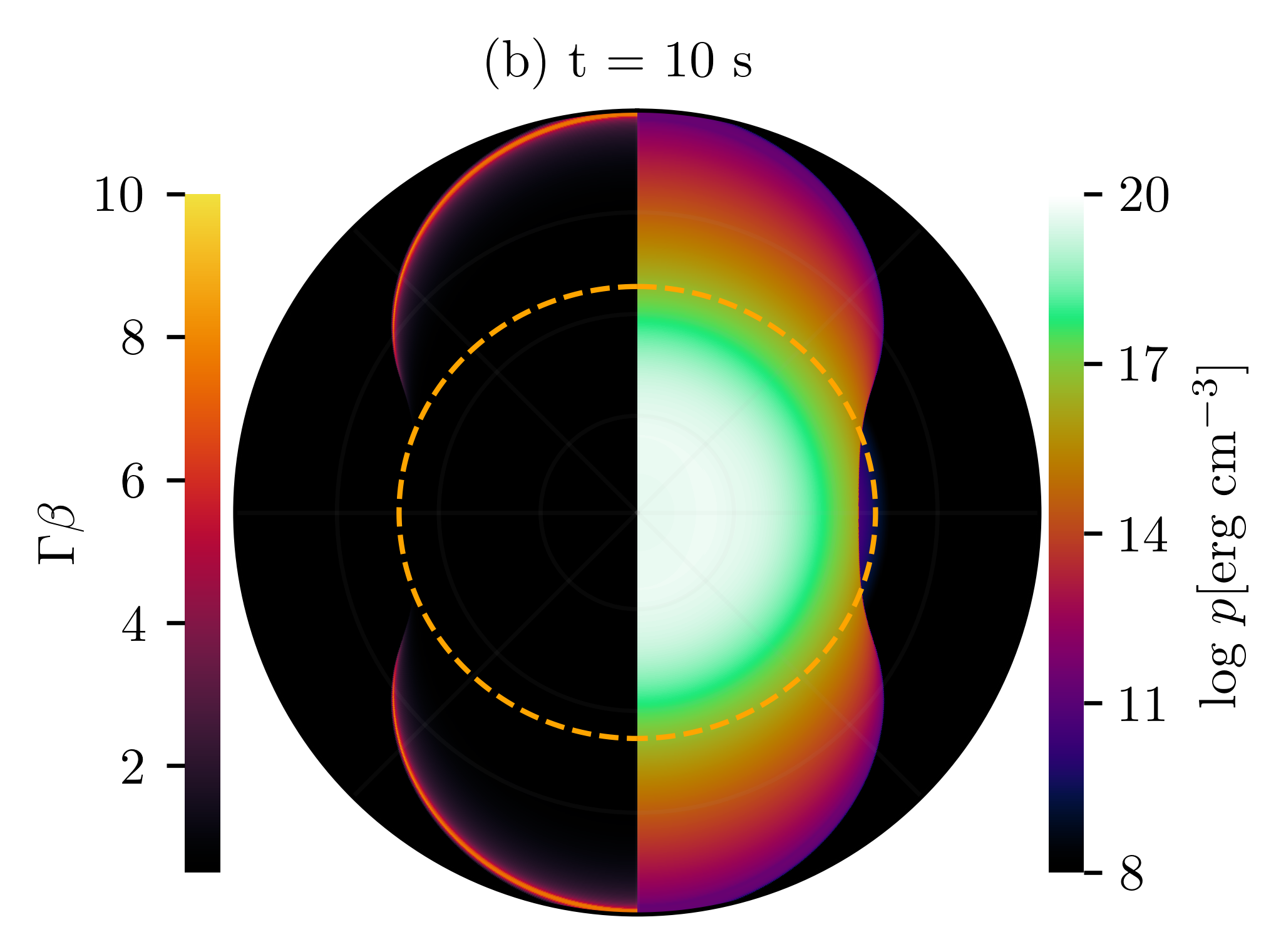}{\columnwidth}{}}\vspace*{-3.5em}
\gridline{\fig{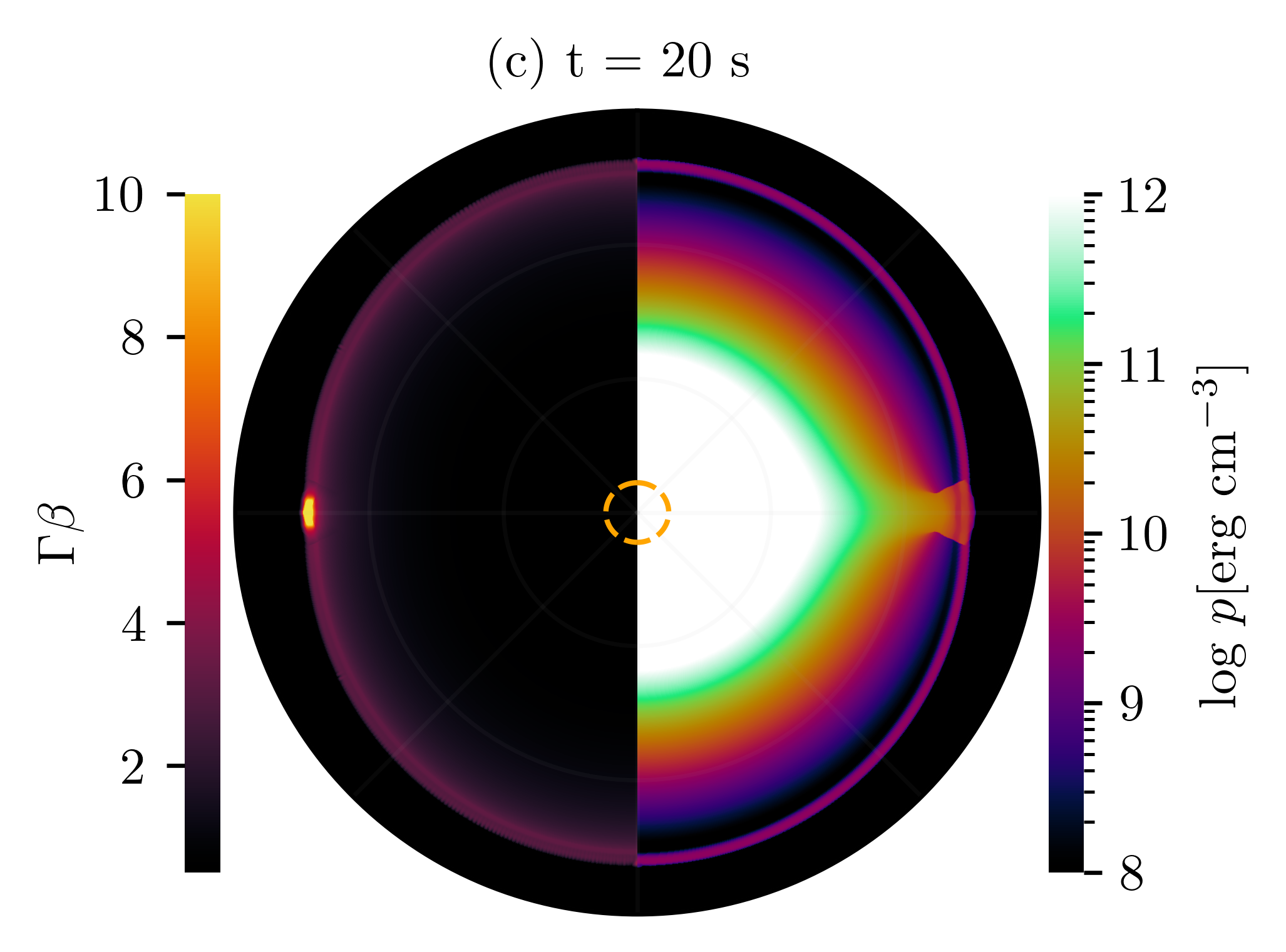}{\columnwidth}{}}\vspace*{-1.0em}
\caption{The dynamical evolution of a $\unit[5 \times 10^{52}]{erg}$ explosion within a 5\% flattened progenitor star (shown as the elliptical dashed line with semi-major axis $\sim\!\unit[4.5 \times 10^{10}]{cm}$.) We first see the deformed prolate shock approach the poles in panel (a). In panel (b), the lobe shocks have spread sideways towards an impending collision point in the equatorial plane. In panel (c), the lobe shocks collide and force material radially outward, forming an expanding relativistic ring.}
\label{fig:shock_evol}
\end{figure}
The simulation domain extends in radius from $r_{\rm in} = 0.01 R_\odot$ to $r_{\rm out} = 200 R_\odot$, and over the whole sphere in latitude, $\theta \in [0, \pi]$. The radial zones are spaced logarithmically and such that $\Delta r = r\Delta \theta$. The explosion radius is $r_{\rm exp} = 1.5 r_{\rm in}$, so the initial high-pressure region is covered by 230 radial zones. We simulate stars with flattening parameters $\varepsilon$ equal to 0.0, 0.05, and 0.2. 

\section{Results} \label{sec: results}
Figure \ref{fig:shock_evol} shows a $\unit[5 \times 10^{52}]{erg}$ point explosion set off inside a star with 5\% flattening, i.e. $\varepsilon = 0.05$. The sequence of images shows the shock evolving first inside the stellar envelope in panel (a), while it is breaking out in panel (b), and post break-out in panel (c). After the shock is first launched inside the star, it develops a prolate geometry, advancing radially faster at the poles than at the equator. This happens because the shock acceleration is proportional to the magnitude of the density gradient \citep[see e.g.][]{Sakurai+1960, Matzner+Mckee+1999}, and the radial density distribution in the flattened star declines faster in the polar direction than in the equatorial plane.

The prolate shock first emerges from the stellar surface at the poles. After it breaks out, the pressurized post-shock region expands freely away from the breakout point, and into the dilute wind medium outside the star, forming lobes in both hemisphere.
The expanding lobes form a peanut shape surrounding the star as shown in panel (b).

The lobe surfaces collide at an oblique angle at the stellar equator, producing a very high pressure behind the point where they intersect. The highly-pressurized region, which we refer to as the \emph{ring}, expands fastest toward the dilute, unshocked gas ahead of the intersection point.

The highly pressurized ring, which was heated by the lobe collision, sometimes reaches ultra-relativistic speeds as it expands radially outwards. Whether it does or not depends on whether the ring can outrun the point of contact between the two lobes; when the lobe collision point moves too fast in the radial direction, the previously shocked-heated ring decelerates as it encounters high-pressure gas that was shocked more recently by the colliding lobes. However, when the lobe intersection point does not move outward too quickly, the high-pressure gas can overtake the lobe intersection point, allowing the ring to accelerate freely into the dilute wind medium and become ultra-relativistic.

The radial speed of the lobe intersection point is proportional to the obliquity of the lobe shocks.\@ If the northern and southern lobe surfaces collide head-on, the intersection point moves rapidly outwards (note that it can be superluminal), whereas, when the lobes intersect at a high obliquity, the intersection point advances not much faster than the shocks themselves.\@ On the other hand, the head-on shock collision produces a higher pressure than the grazing collision does.\@ This picture suggests that an optimal lobe geometry should exist, for which the shock collision is sufficiently head-on to produce extreme heating, but also sufficiently grazing that the shock-heated gas can outrun the lobe intersection point and accelerate freely in the dilute wind medium. Since the lobe geometry is determined by the flattening parameter $\varepsilon$, there should then be a critical value of $\varepsilon$ which maximizes the amount of relativistic material launched in the explosion.

\begin{figure}
\centering
    \includegraphics{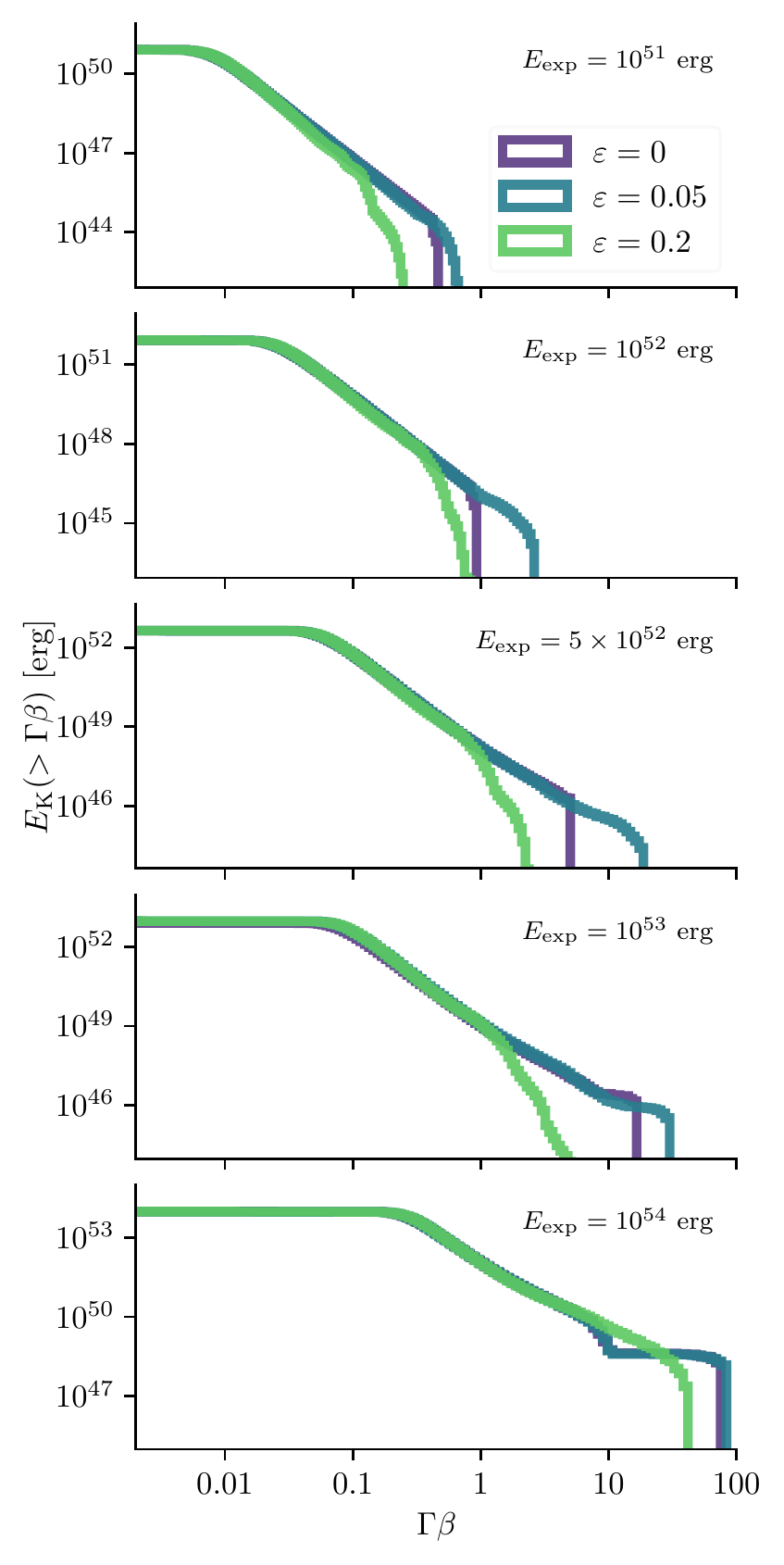}
    \caption{The cumulative distributions of kinetic energy, $E_k(>\Gamma \beta)$, for a suite of explosions with energies between $\unit[10^{51}]{erg}$ and $\unit[10^{54}]{erg}$ and oblateness magnitudes $\varepsilon=$ 0, 0.05 and 0.2. These distributions were recorded at times when the gas is approximately ballistic, from top to bottom at times 93s, 35s, 22s, 12s, and 20s post explosion. For all but the largest explosion energy, stars with 5\% flattening ($\varepsilon = 0.05$) produce faster ejecta than their spherical counterparts. In general, the spherical star $\varepsilon=0$ and the over-flattened star $\varepsilon=0.2$ produces less relativistic material than the optimally flattened star ($\varepsilon=0.05$). 
    \label{fig:suite}}
\end{figure}

This expectation is consistent with the measured distribution of kinetic energy $E_k(>\Gamma \beta)$ over gas four-velocity $\Gamma \beta$, shown in Figure \ref{fig:suite} for a range of different explosion energies and flattening parameters. A total of 15 simulations are shown in that figure: three simulations, with $\varepsilon$ values of 0.0, 0.05, and 0.2, are run for each of five explosion energies: $10^{51}$, $10^{52}$, $5 \times 10^{52}$, $10^{53}$, and $10^{54}$ erg. More simulations were performed, but these values of the oblateness capture the main features of the explosion dynamics. We find that the stars with 5\% flattening consistently accelerate more relativistic material than a spherical star, in some cases by a significant margin. For explosion energies in the range $\unit[10^{51}-10^{53}]{erg}$, the cutoff of the kinetic energy distribution extends to higher speeds, by factors of several for $\varepsilon = 0.05$ as compared to the spherical cases. For example, in the simulations where $E_{\rm exp}$ is $\unit[5 \times 10^{52}]{erg}$ and $\unit[10^{53}]{erg}$, the optimally flattened star has $\sim \unit[10^{46}]{erg}$ of kinetic energy above $\Gamma \beta = 10$ and $\Gamma \beta = 20$, respectively, whereas the equivalent explosions in the spherical star had no energy in such fast material. Importantly, we also find that excessive flattening can hinder the acceleration of relativistic material. For example, as can be seen in Figure \ref{fig:suite}, simulations with $\varepsilon = 0.2$ produce less relativistic material than the spherical and optimally flattened cases.

\begin{figure}
\centering
\includegraphics{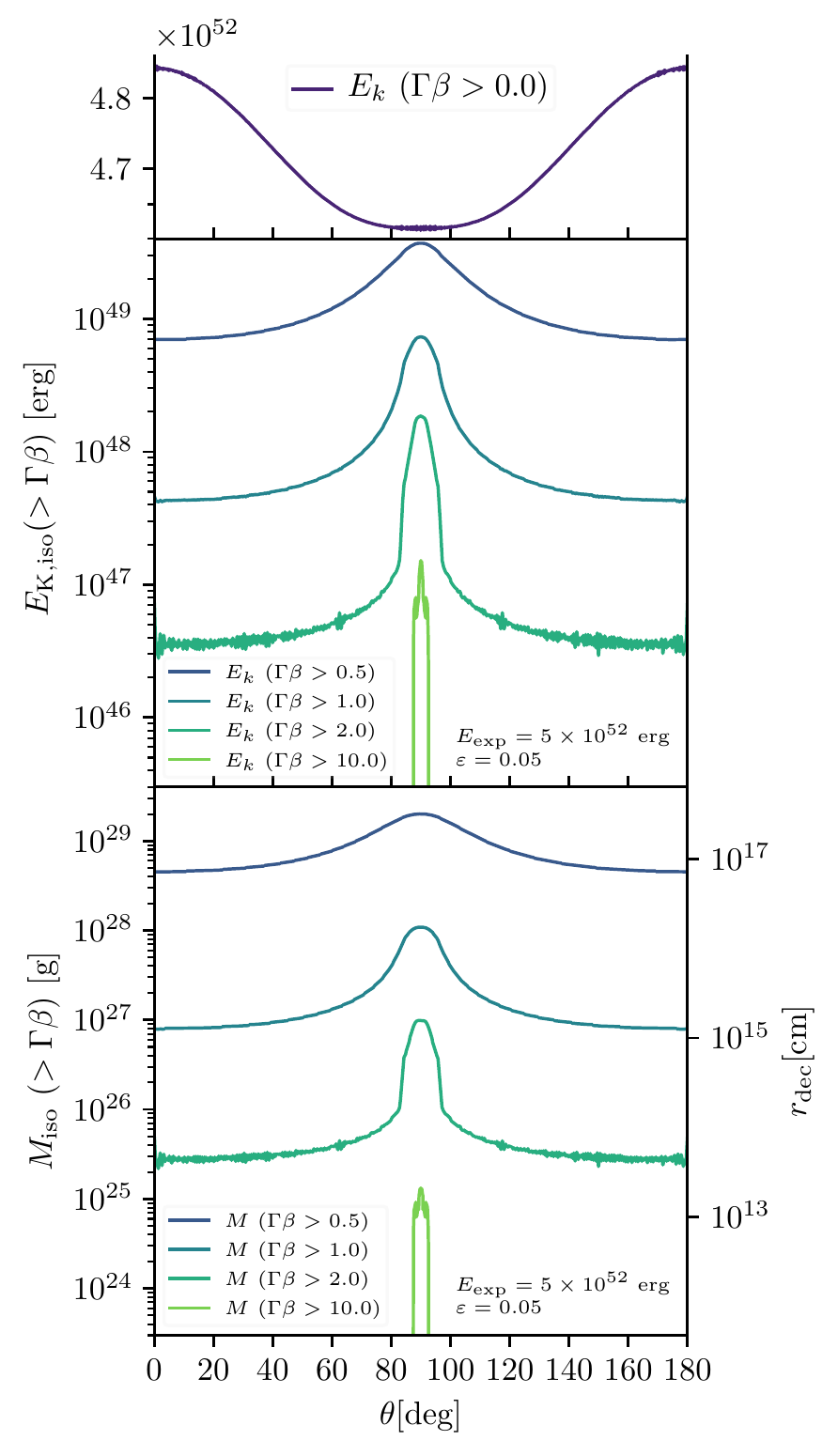}
\caption{\emph{Top}: The cumulative distribution of isotropic-equivalent kinetic energy $E_{k, \rm{iso}}(>\Gamma \beta, \theta)$, as a function of the polar angle $\theta$, for representative values of $\Gamma \beta$. The small axes at the top show $E_{k, \rm{iso}}(>0, \theta)$, i.e. the total isotropic-equivalent kinetic energy as a function of polar angle. The total energy in the domain, above each of the $\Gamma\beta$ cutoffs $0.5$, $1$, $2$, and $10$, is $\unit[1.6 \times 10^{49}]{erg}$, $\unit[1.75 \times 10^{48}]{erg}$, $\unit[1.8 \times 10^{47}]{erg}$, and $\unit[3.0 \times 10^{45}]{erg}$, respectively. \emph{Bottom}: The cumulative distribution of isotropic-equivalent mass $M_{\rm iso}(>\Gamma \beta, \theta)$, as a function of $\theta$, for representative values of $\Gamma \beta$. The secondary vertical axis at the right shows the nominal deceleration radius $r_{\rm dec}$ (Equation \ref{eq: dec_rad}) corresponding to the mass scale at the left, when $A_*$ has the fiducial value of $0.1$.
\label{fig:var_per_theta}}
\end{figure}

The top panel of Figure \ref{fig:var_per_theta} shows the isotropic-equivalent kinetic energy, $E_{k, \rm iso}(>\Gamma \beta) \equiv 4 \pi \times dE_k(>\Gamma \beta)/d\Omega$, as a function of polar angle $\theta$ for representative values of $\Gamma \beta$, for a $\unit[5 \times 10^{52}]{erg}$ explosion. We see that the kinetic energy in fast material, $E_{k,\rm iso}(\Gamma\beta > 1)$, is concentrated around the equator, being at least a factor of 10 higher than at the equator for the $\Gamma \beta > 1$ material. In contrast, the total kinetic energy, $E_{k,\rm iso}(\Gamma\beta > 0)$, is slightly more concentrated at \emph{high} latitudes than at the equator. These differences in the latitudinal energy distribution of fast and slow material may have implications for the evolution of the remnant morphology, which we discuss in Sec. \ref{sec: discussion}. The angular extent of the relativistic ring can be measured from the cutoffs of the $E_{k,\rm iso}(\Gamma\beta > 1)$ shown in Figure \ref{fig:var_per_theta}; we find the ring subtends $\Delta \theta \simeq 6^\circ$ in latitude around the equator. This corresponds to roughly 5\% of the sphere.

The bottom panel of Figure \ref{fig:var_per_theta} shows the isotropic-equivalent mass, $M_{\rm iso}(>\Gamma \beta) \equiv 4 \pi \times dM(>\Gamma \beta)/d\Omega$, as a function of $\theta$ for the same simulation model (explosion energy $\unit[5 \times 10^{52}]{erg}$, flattening $\varepsilon = 0.05$).

Based on the quantity of mass launched at each latitude, and the density of the ambient wind-like medium, we can compute a nominal deceleration length scale,
\begin{equation}
    \label{eq: dec_rad}
    r_{\rm dec}(\Gamma \beta, \theta) \equiv \frac{M_{\rm iso}(>\Gamma \beta, \theta)}{4\pi A},
\end{equation}
which is the radius at which the coasting shell sweeps up mass in the ambient medium comparable to its own, i.e., $M_{\rm ambient} = M_{\rm iso}(>\Gamma \beta, \theta)$. Note that this deceleration radius is an approximation for the bulk flow and would be smaller for the more relativistic fluid parcels by a factor $1 / \Gamma_{\rm ej}$. We've also ignored effects due to spreading of the ring. We find that $M_{\rm iso}(\Gamma \beta > 1, \theta=90^\circ) \simeq 6 \times \unit[10^{27}]{g} \simeq M_\oplus$. The secondary vertical axis at the right side of the lower panel of Figure \ref{fig:var_per_theta} shows the nominal deceleration radius corresponding to the mass of ejecta on the left vertical axis. We use the nominal deceleration length scale of the different ejecta components in Sec. \ref{sec: discussion} to speculate about the long-term evolution of the spatial distribution of the explosion ejecta in the remnant of a ring-like explosion.

\section{Astrophysical Implications}
\begin{figure*}
    \centering
    \includegraphics{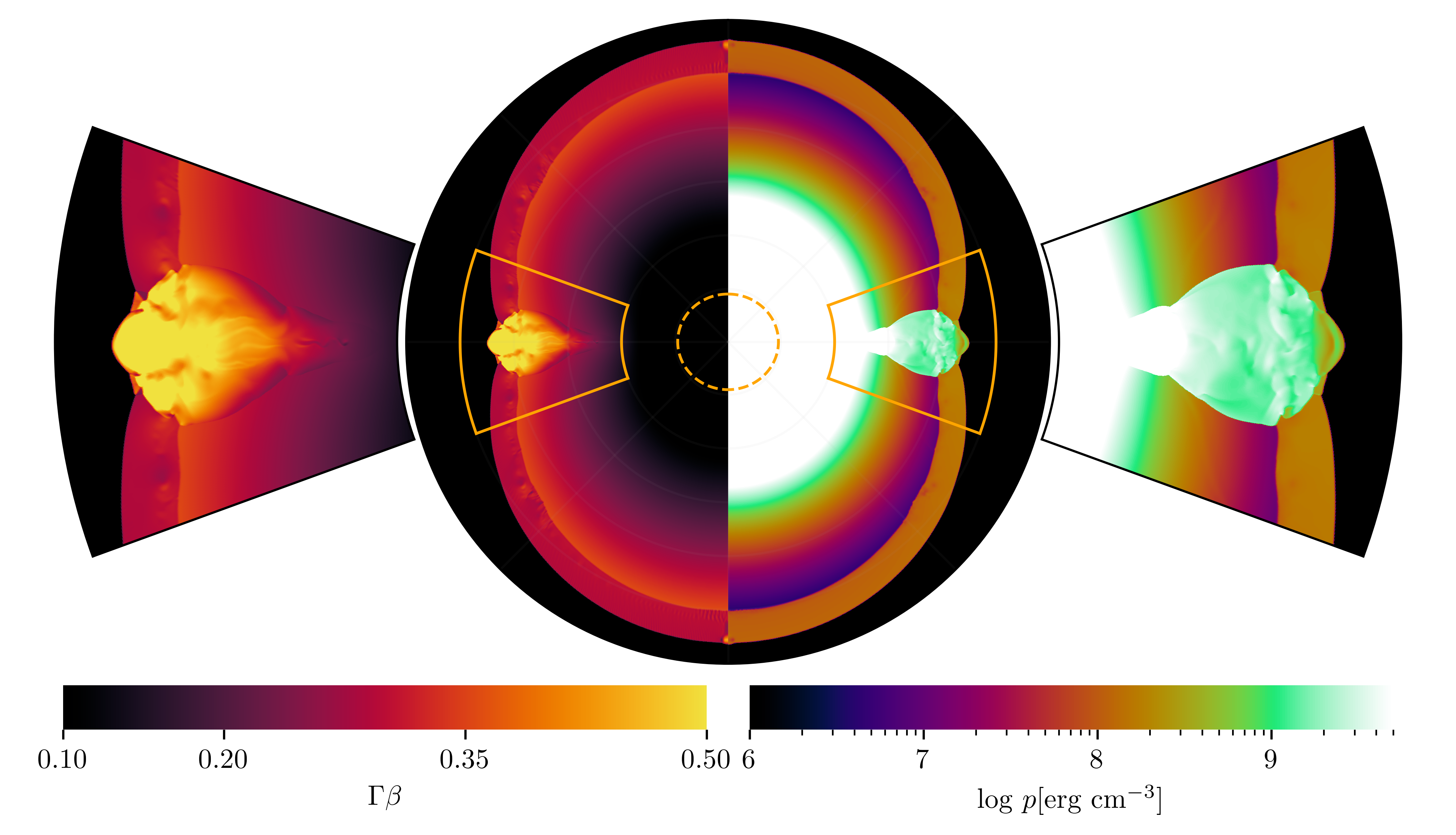}
    \caption{The pressure and 4-velocity at $t = \unit[93]{s}$ showing a cross-section ($r$--$\theta$ slice) of the explosion dynamics. The expanding mildly relativistic ring, seen also in the expanded wings, is produced at the equator by the collision of expanding lobes propagating from the poles during a $E_{\rm exp} = \unit[10^{51}]{erg}$ point explosion in an oblate star with $\varepsilon = 0.05$. The dashed orange ellipse shows the original size and shape of the progenitor. In the right wing, we see a steep pressure gradient point in the $+\hat{r}$ direction which is the main driver behind the equatorial flow. The left wing shows that the ring expands radially outward with mildly relativistic speed of $\Gamma \beta \sim 0.5$.}
    \label{fig:ring}
\end{figure*}
\label{sec: discussion}
%


\subsection{Remnant morphology}
The morphology of an ellipsar remnant starts out ring-like, and is expected to become more spherical over time. The time for the remnant to lose its ring-like morphology is inversely proportional to the density of the circumburst medium.
Using Eq. \ref{eq: dec_rad} and our result from explosion energy $\unit[5 \times 10^{52}]{erg}$, we estimate the deceleration time of an ejecta parcel to be
\begin{eqnarray}
    \label{eq: dec_time}
    t_{\rm dec} &=& \frac{r_{\rm dec}}{\beta c} \\
    &\simeq& 247 \left(\frac{M_{\rm ej}}{\unit[2\times 10^{29}]{g}} \right) \left( \frac{A_*}{0.1}\right)^{-1}\left(\frac{\beta}{0.5} \right)^{-1} \rm{day} \, , \nonumber
\end{eqnarray}
where the reference values above were evaluated at the equator, and for the mildly relativistic ejecta $\Gamma \beta = 0.5$. The mildly relativistic mass ejected along the poles is $3-4$ times lower; $t_{\rm dec,pole}(\Gamma \beta >0.5)\big|_{\theta = 0} \simeq 54$ days.\@ The deceleration time difference between the equatorial flow and the polar flow indicates that the expanding remnant should maintain its ring-like morphology for nearly a year. We note that the nominal deceleration time is a good approximation for the mildly relativistic ejecta, but for the more relativistic fluid parcels, the deceleration time scale is shorter by a factor $1 / \Gamma_{\rm ej}$. Since the slow ejecta containing most of the mass expand slightly faster in the polar direction than the equatorial direction, the overall late-time morphology of the ellipsar remnant is expected to be mildly prolate. 

\subsection{Afterglow and Polarization}

Due to the ring-like morphology of the relativistic ejecta from ellipsars, the external shock that it forms as it sweeps up the surrounding medium is novel and will produce an afterglow light curve that is distinct from a jet-driven afterglow, as will be presented in a future publication. The fastest ejecta is concentrated within $6^\circ$ of the equator and beams its radiation to a great circle covering $5\%$ of the sky, whereas bipolar jets with opening angle $\theta_j = 6^\circ$ beam their radiation to only about $\theta_j^2 / 2 = 0.5\%$ of the sky. The required rate of ellipsars to account for a given observed population of transients is thus significantly lower than the required rate for a jet-like explosion.

Ellipsar afterglows may be polarized, depending on viewing angle, perhaps analogously to the polarization effects discussed by, e.g., \citet[]{Gruzinov+Waxman+1999} for jet-driven afterglows.\@ In addition, we expect the SNe light from ellipsars to be mildly linearly polarized since it originates in the photosphere of the slower stellar material which expands as a mildly prolate spheroid.


\subsection{Was SN1998bw/GRB980425 an ellipsar?}

The first observed connection between supernovae and GRBs was reported by \citet[]{Galama+1998} who discovered SN 1998bw in the error box of the low luminosity GRB 980425.\@ GRB 980425 was the first of a set of GRBs with a $\gamma$-ray energy thousands to millions times smaller than that of classical GRBs. Subsequently, a few other GRBs with similar characteristics to GRB 980425 --- low-redshift, small Lorentz factor, and low-luminosity --- were observed which helped classify GRB 980425 as part of a distinct GRB population termed llGRBs \citep[][]{Liang+2007,Virgili+2009}.\@ \citet[]{Tan+2001} explored the possibility that GRB980425 was produced by shock acceleration during breakout of an energetic explosion from a stripped envelope star. We compare our results with those of \citet{Tan+2001} who used a less massive progenitor for their simulations such that $E_{\rm exp}/E_{\rm rest} = 3.3 \times 10^{-3}$, which is about twice the value we use for our progenitor-explosion configuration. They conclude that a spherical, compact carbon-oxygen core of mass $5 M_\odot$ and explosion energy a few$\times 10^{52}$ erg is sufficient for producing the energetics of GRB 980425. Figure \ref{fig:suite} shows that all of our $5 \times 10^{52}$ erg explosions in the spherical and flattened stars produce enough kinetic energy in $\Gamma \beta > 1$ ejecta to power GRB 980425 in rough agreement with \citet{Tan+2001} (even though our progenitor mass of $18 M_{\odot}$ is 3.6 times larger), but with an interesting caveat. Namely, point explosions in ``ideally'' flattened ($\varepsilon \lesssim 0.05$) stars tend to focus their fastest ejecta into an equatorial ring with higher outflow speed than their spherical counterparts. We find that for ejecta speeds $\Gamma \beta > 1$, there is, on average, an order of magnitude more energy at the equator than the poles for flattened stars. We also note that our model with $5 \times \unit[10^{52}]{erg}$ and {\eps} = 0.05 was capable of producing Lorentz factors $\Gamma > 10$, approaching the upper boundary for llGRBS \citep{Liang+2007}. For the same explosion energy, but  with {\eps} = 0, no such highly relativistic material was present in both our simulations as well as the simulations presented in \citet[]{Tan+2001}.




\subsection{Event rates from radiating relativistic rings}

We report the isotropic equivalent kinetic energy as a function of polar angle in Figure \ref{fig:var_per_theta}. Assuming the radiation of the $\Gamma \beta > 10$ material is strongly beamed in the forward direction, then the isotropic equivalent kinetic energy at the equator is $E_{k, \rm iso}(>10)\big|_{\theta = 90^\circ} \sim \unit[10^{47}]{erg}$. Note that this isotropic equivalent energy is only valid for high beaming factors. At the more modest beaming factors below $\Gamma \sim 10$, the radiation is more or less isotropic, so the effective energies for $\Gamma \beta < 10$ come down by a factor of $4 \pi$. For example, the true inferred energy for $\Gamma \beta > 1$ at the equator is roughly $\unit[6 \times 10^{47}]{erg}$ and not $\unit[8 \times 10^{48}]{erg}$ as Figure \ref{fig:var_per_theta} panel (a) might suggest. Furthermore, we find that the transient is geometry-dependent like a jet, but with a viewing probability
%
\begin{equation}
    P(\Delta\theta) = \frac{1}{4\pi} \int_{\pi/2 - \Delta\theta/2}^{\pi/2 + \Delta\theta/2} d\Omega \simeq \frac{\Delta \theta}{2}.
\end{equation}
This has the following observational implications: (1) the occurrence rates of these events are higher, but the detection rates might be comparable due to the relatively low luminosity; (2) at best, we observe the ring edge-on and receive photons from at least half of the ring until it spreads open perpendicular to our light of sight, synonymous to that of a clam shell, or at worst, we are perpendicular to said ring and receive photons some time later once the emitting ring has slowed down enough. Only at early times if viewed edge-on will this event be seen as a relativistic equatorial outflow. 

\subsection{Fast Blue Optical Transients}

Ellipsars with SNe-like explosion energies ($\sim\unit[10^{51}]{erg}$) can produce mildly relativistic explosions, similar to the speeds inferred from observations of FBOTs \citep[see][]{Drout-2014, Margutti-2019, Coppejans-2020, Perley-2021}. These transients have shown strong polar-equatorial velocity gradients in their spectra, suggesting an aspherical explosion; and fast ejecta with $v>0.1c$ is inferred due to the presence of broad absorption lines. It has recently been proposed that relativistic jets powered by a long-lived engine may be responsible for producing the fast ejecta \citep{Soker+2019, Gottlieb2022, Soker2022}.\@ Our results indicate that the ellipsar mechanism may be capable of producing similar mildly relativistic ejecta but without requiring the formation of a jet. Figure \ref{fig:ring} shows the formation of a mildly relativistic ring with $\Gamma \beta \lesssim 0.5$ resulting from a $\unit[10^{51}]{erg}$ explosion, though it carries less mass than inferred for some FBOTs. 

\subsection{High energy cosmic rays}
If a significant fraction of massive stars are mildly oblate and explode as ellipsars in low density environments, then in the Galactic population of stellar explosions the fraction of explosion kinetic energy that is trans-relativistic is significantly increased.\@ Ellipsars could thus be a primary source of high energy ($\unit[10^{15}-10^{18}]{eV}$) cosmic rays due to their production of trans-relativistic  ejecta \citep{Budnik+2008}.

\section{Summary and conclusions}
This work has demonstrated the following: (1) point-like explosions in oblate stars with an optimal flattening of $\varepsilon \simeq 0.05$ produce more relativistic ejecta than similar explosions in spherical stars; (2) overly flattened stars with $\varepsilon \gtrsim 0.2$ diminish the production of relativistic ejecta; (3) the relativistic ejecta are focused in a ring, which expands outwards in the equatorial plane of the progenitor; (4) explosions of oblate stars might produce transients including trans-relativistic supernovae and Fast Blue Optical Transients (FBOTs), as well as low-luminosity gamma-ray bursts (llGRBs), if viewed in the equatorial plane; (5) Supernova-like ($\gtrsim \unit[10^{51}]{erg}$) explosions in progenitors with $\varepsilon \lesssim 0.2$ and embedded in sufficiently dilute environments (mass loss rates $\lesssim \unit[10^{-5}]{M_\odot yr^{-1}}$ for a $\unit[10^3]{km \ s^{-1}}$ wind) are capable of accelerating mildly relativistic ejecta.

\begin{acknowledgements}
MD thanks Andrei Gruzinov for many helpful discussions and acknowledges a graduate fellowship at the Kavli Institute for Theoretical Physics during which this research was completed. AM acknowledges support from NSF grant AST-1715356. The Python package \texttt{CMasher}, a color-vision-deficiency-friendly scientific colormap extension to matplotlib \citep{CMasher} was used in the creation of plots.
\end{acknowledgements}




\end{document}